# RF heating experiments with a TESLA-9-cell cavity towards in-situ low- / mid-T-baking

H.-W. Glock[†], J. Knobloch[1], J.-M. Köszegi, A. Velez[2],

Helmholtz-Zentrum Berlin für Materialien und Energie, Berlin, Germany

[1]also at University of Siegen, Siegen, Germany
[2]also at Technical University Dortmund, Dortmund, Germany

*Abstract*

Under-vacuum low- and mid-temperature baking revealed beneficial effects on the performance of niobium-made cavities for superconducting radio-frequency (SRF) applications, primarily seen in particle accelerators. Such a baking process is typically performed in a dedicated oven. In this paper the experimental investigation is described, whether an appropriate heating of an elliptical 9-cell 1.3 GHz TESLA cavity is feasible using rf power, which would be a pre-condition for a processing done fully in-situ without the need of costly and risky dismantling / remounting operations. It is demonstrated that such a heating is possible, whilst complicated by uneven heating rates in the individual cavity cells.

## I. INTRODUCTION

Under-vacuum baking of niobium radio-frequency (rf) cavities with low (about 120 °C) or mid-range temperatures (200 °C to 350 °C) demonstrated beneficial effects on their performance under superconducting conditions, in particular their resonance quality factor (e.g. [1], [2] and references therein). Such so-called superconducting radio-frequency (SRF) cavities are mainly used in particle accelerators in order to provide high accelerating field strengths with very low dissipative energy loss in the cavity walls. Such cavities need to be installed in housing structures, denoted as modules, which provide mechanical support, supply of liquid He cooling, magnetic shielding and thermal insulation by evacuation and both actively cooled and passive thermal shields. Dismantling or re-mounting an SRF cavity from or into such a module is technologically demanding and implies high efforts together with severe risks of performance degradations due to particulate contaminations. Therefore it would be highly desirable to perform baking processes in-situ of the module, preserving the evacuated status of the cavity and keeping all its connections untouched. It would furthermore be highly effective and elegant to heat the cavity in its normal conducting status using rf power to be dissipated in the cavity's wall being fed into the cavity using the same coupler and rf line installations foreseen for the superconducting cavity operation.

This paper describes a first in-vacuum rf heating experiment with a niobium made TESLA-9-cell SRF cavity [3] which aimed for getting initial hands-on experiences with the process and the identification of potential difficulties and drawbacks. A preparatory experiment in air was executed in advance, which is described in [4]. Therefore the setup featured only the most essential components which were – apart from the cavity – a manually pre-tuneable on-axis antenna mounted at one of the cavity's beam pipe flanges, a pick-up antenna at the opposite end and a set of thermal sensors adhesively connected to the outer side of each cell and both beam pipes (plus additional spots in the cryostat). The system was housed in an isolating vacuum inside HZB's cavity test cryostat HoBiCat [5], which itself is operated inside a radiation protection bunker. HoBiCat is typically used for experiments at cryogenic temperatures, but it is also compatible with elevated temperatures. Rf power was supplied from the outside of the bunker using a broadband amplifier of (nominally) 300 W power and a protective circulator. The amplifier was initially driven by a dedicated frequency generator, which was replaced in the course of the experiments by a direct connection to a vector network analyzer (VNA) output, allowing to directly monitor the pick-up return signal within a single device. Temperature and rf power data also were automatically stored in HoBiCat's data archiving system. No additional infrared radiation shield between the cavity and the cryostat wall was installed.

The experiments reported here were conducted in four different days with sufficient time in between to cool down the set-up to similar initial conditions. Main attention was given to the heating rates of the individual cells depending on the input signals's frequency. Those turned out to be rather inhomogeneous and strongly temperature dependent. This was conceptionally expected because of the cell's individual thermal expansions, which depend on the actual (and previous) heating. That in turn depends on the square of each cell's field strength, i.e. the cavity's field flatness profile. The amount of that effect was so significant, that the application of (almost) the full fundamental passband mode set was studied as a possible remedy against uneven heating.

* work supported by grants of the Helmholtz Association
† hans.glock@helmholtz-berlin.de

The results of those experiments are summarized in Chapter III whilst in Chapter II pictures and details about the setup are given. In Chapter IV conclusions are summarized and an outlook about experimental improvements under preparation is given.

## II. EXPERIMENTAL SETUP

The core component of the experiments described here was a RRR300 1.3 GHz 9-cell cavity produced by ACCEL Instruments in 2003/2004 [5] (cf. Fig. 1). The cavity had no enclosing helium vessel. Both HOM couplers were mounted but not used here. The beam pipe port at the pick-up side and the fundamental power coupler port remained open, which allowed evacuation of the inner cavity volume together with the cryostat's isolation vacuum (better 10^-3 mBar). In order to reduce the rf reflection at the power input, a manually tuneable on-axis antenna was chosen as a provisional power coupler, which was connected to the FPC-side beam pipe port of the cavity using a coaxial line segment of adjustable length (cf. Fig. 1, right). The antenna length was tuned before closing the cryostat at room temperature conditions to get an rf match better than 8 dB for the (accelerating) π-mode and also better than 12 dB for the modes $5\pi/9$, $6\pi/9$ and $7\pi/9$. Once installed in the cryostat the coupling was not adjustable any longer. Previous electromagnetic field simulations starting with a well-tuned cavity shape revealed a major influence of the antenna on the π-mode's flatness, reducing the electrical peak field in the antenna side end cell by 28%, compared to the opposite end [ST2024]. A similar influence needed to be assumed in practice, even though no field profile measurements were done in advance and the tuning status of the cavity was unknown. An uncalibrated pick-up antenna was mounted at the cavity's pick-up port located at the beam pipe opposite to the one equipped with the main power antenna.

The test cavity was previously used for display purposes for several years while kept open and exposed against atmosphere. This was acceptable for such a kind of pure warm rf experiments not probing the cryogenic performance.

The cell temperatures, which are the primary quantities of interest in this experiments, were measured using PT100 sensors ("Pt100", TESLA BLATNÁ, Blatná, Czech Republic), specified up to 400°C, attached to the outer cell surface on the so-called equator weld. A proper thermal contact was provided by a commercial silicon-based heat paste ("WLP4", Fischer Elektronik GmbH, Lüdenscheid, Germany), useable up to 250°C, mechanical connection by the use of adhesive Kapton© tape. Four-wire connections were installed linking the sensors with the read-out units ("Model 224", Lake Shore Cryotronics Inc., Westerville, OH, USA), but without individual calibration of the sensors' temperature characteristics. This gave explanation of slightly varying initial temperature readouts. The sensor attached to cell 5 failed initially, the one attached to cell 2 during the second day of operation. Further sensors were installed at the support structure and the cryostat's outer wall, but no relevant results need reporting.

All temperatures and the rf power data (forward, reflected and transmitted through the pick-up) were stored automatically, whilst the transmitter/VNA frequency settings were kept manually.

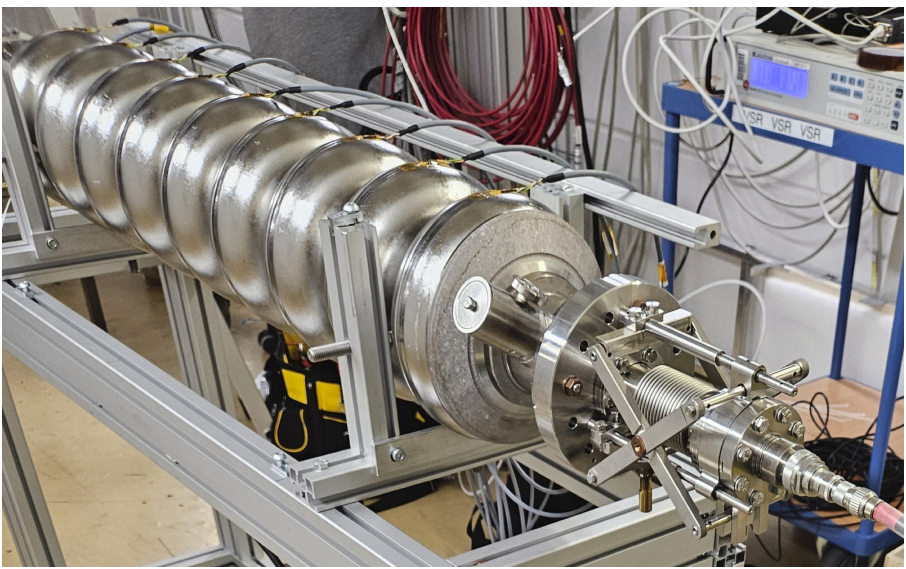
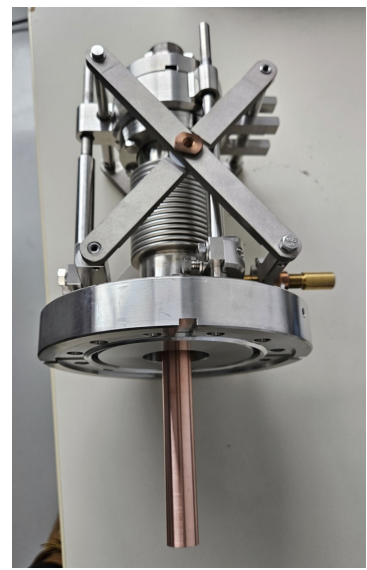

Figure 1: (left) Niobium TESLA-9-cell cavity with temperature sensors on top of all cell equators and adjustable antenna (separately shown right) mounted at the right-hand side beam pipe before installing in the HoBiCat cryostat.

The rf power was supplied by a broadband power amplifier ("BBA130-D300", Rohde&Schwarz, Munich, Germany), driven (alternatively) by a cw rf signal generator ("SML03", Rohde&Schwarz) or an VNA ("E8358A", Agilent). In both cases a circulator ("VALVO VAN1053A", Microwave Techniques GmbH, Hamburg, Germany) was used to prevent amplifier damages caused by reflected power. Cable and circulator losses between transmitter and antenna port were determined to – 2.7 dB, which corresponds to 180 W of incident power when the transmitter delivered 335 W as observed during the experiments.

### III. EXPERIMENTAL OBSERVATIONS

Already the very initial experiment shown in Figure 2 illustrated, beside the need of a proper frequency setting, a strong self-detuning of the cavity caused by thermal expansion (cf. the decrease of the pick-up signal strength by about -5 dB in the time interval $9 \cdot 10^3$ s to $13 \cdot 10^3$ s). Secondly a very inhomogeneous heating (strongly pronounced in cells 3 and 4, moderate in cells 1, 2 and 8, very low in cells 6, 7 and 9, defect sensor in cell 5) was observed once the tuning was adjusted based on the pick-up signal strength (time after $13.8 \cdot 10^3$ s). In average every ~ 500 s a re-tuning was needed in order to keep the amplitude loss less than ~ 3 dB. Certain frequency adjustments caused clear changes in the heating rates of individual cells (e.g. increase in cell 2 and decrease in cell 8 after $14.9 \cdot 10^3$ s).

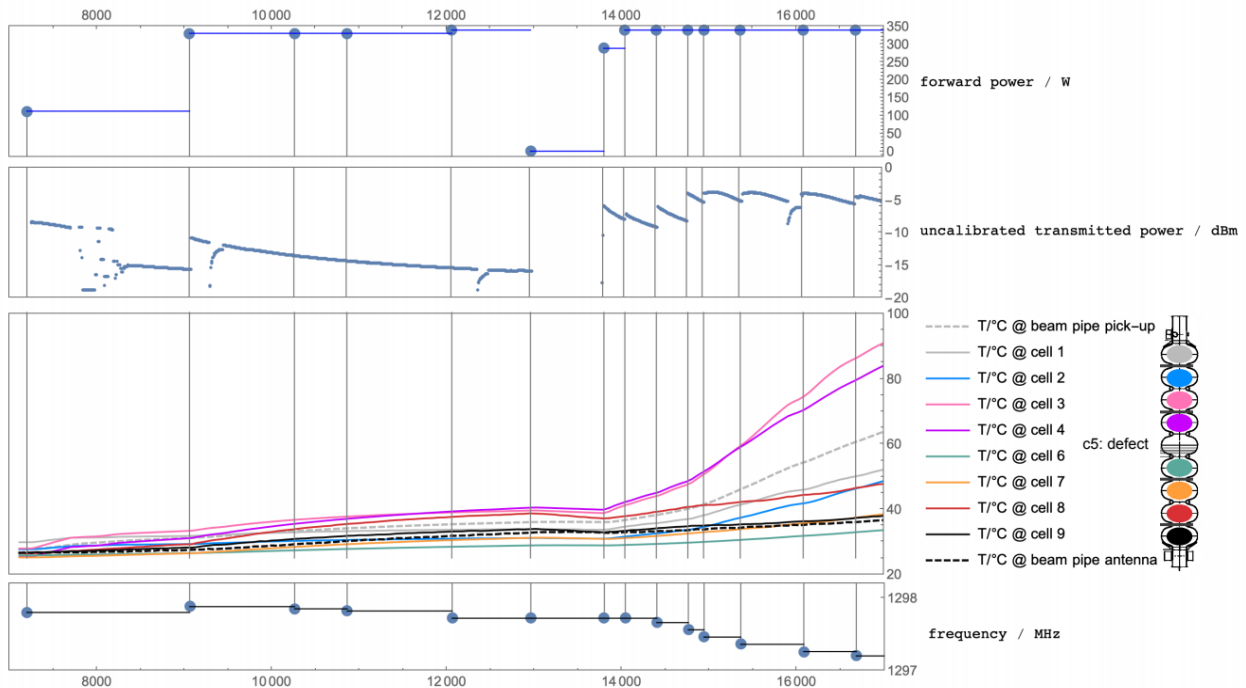

Figure 2: First heating results (all graphs depending on the same time scale, given in seconds, blue dots and vertical lines indicating change of set values) *bottom to top:* transmitter frequency setting (MHz); temperature measurements in all cells and both beam pipe extensions (°C, sensor at cell 5 defect); (uncalibrated) measured pick-up power (dBm); amplifier power set values (W).

Because of the observed inhomogeneity of heating rates, the use of the $5\pi/9$-mode was tested in the further course of the initial experiment, as shown in Fig. 3 after $t = 17.5 \cdot 10^3$ s. This clearly reduced the heating rates in cells 1 to 4, which had the fastest raise before, whilst somewhat accelerated heating in cells 6 to 8. Nevertheless, the temperatures in the latter ones still remained clearly lower than in the hottest cells 3 and 4.

Cell 7 showed a remarkable self-tuning behaviour around $t \sim 19.5 \cdot 10^3$ s, when its heating rate accelerated strongly without the change of any parameters; this saturated again after about 500 s. A similar process happened in cell 1 almost simultaneously but with a lower temperature gain.

From Fig. 3 also a very low mutual thermal coupling via heat conduction or radiation can be derived in view of the significant temperature differences that remained between neighbouring cells (e.g. cells 2 and 3, 6 and 7) during the entire observed time span. A small thermal coupling is indicated by the temperature maximum of cells 4 and 8 after the power shut down at $t \sim 20.7 \cdot 10^3$ s, which is delayed by a short time span during which those cells experience a small additional heating from their warmer neighbours (cells 3 and 7 resp.).

Further experiments with passband modes lower than 5π/9 revealed no beneficial effects on the heating performance, in particular since the antenna matching went worse to lower frequencies.

Replacing the signal generator by a VNA and reading the transmission through the pick-up antenna within a single device facilitated the observation of thermal detuning and re-adjusting of the driving frequency band, using slow (100 s) and narrow (variable between 20 kHz to 200 kHz) frequency sweeps. (The sweeping frequency is reflected in the bar representation in the frequency plots of Fig. 4 instead of lines in Figs. 2 and 3.)

Figure 4 shows the result of the longest heating experiment executed so far, covering ~ 6.7 hours of active rf power, using the four modes 5π/9, 6π/9, 7π/9 (predominantly) and π. Whilst the 5π/9- and 6π/9-mode, which were applied in the beginning, delivered very similar heating patterns, changing to 7π/9 (at $t \sim 195 \cdot 10^3$ s) caused a significant deceleration in cells 3 and 7, initially with an even negative heating rate. Heating in cell 6 was accelerated for a certain time (around $t \sim 198 \cdot 10^3$ s) and overtook cells 7 and 8 before getting back to an average slope. Test-wise switching to the π-mode ($t \sim 204 \cdot 10^3$ s) severely decelerated all cells except of cells 3 and 1, the latter also showing in total three short temperature excursions, which are in lack of explanation. Therefore tuning was reset soon to the 7π/9-mode, which resulted in a slowing further temperature increase in all cells except of cell 6. The latter even began to cool down ($t \sim 209 \cdot 10^3$ s), which then gave reason to terminate the experiment ($t \sim 213 \cdot 10^3$ s). Finally cell 4 arrived at the highest temperature of 150 °C, which was 115 °C above starting value, whilst the coldest cell 9 only reached 70 °C, which is 40 °C above the initial readout.

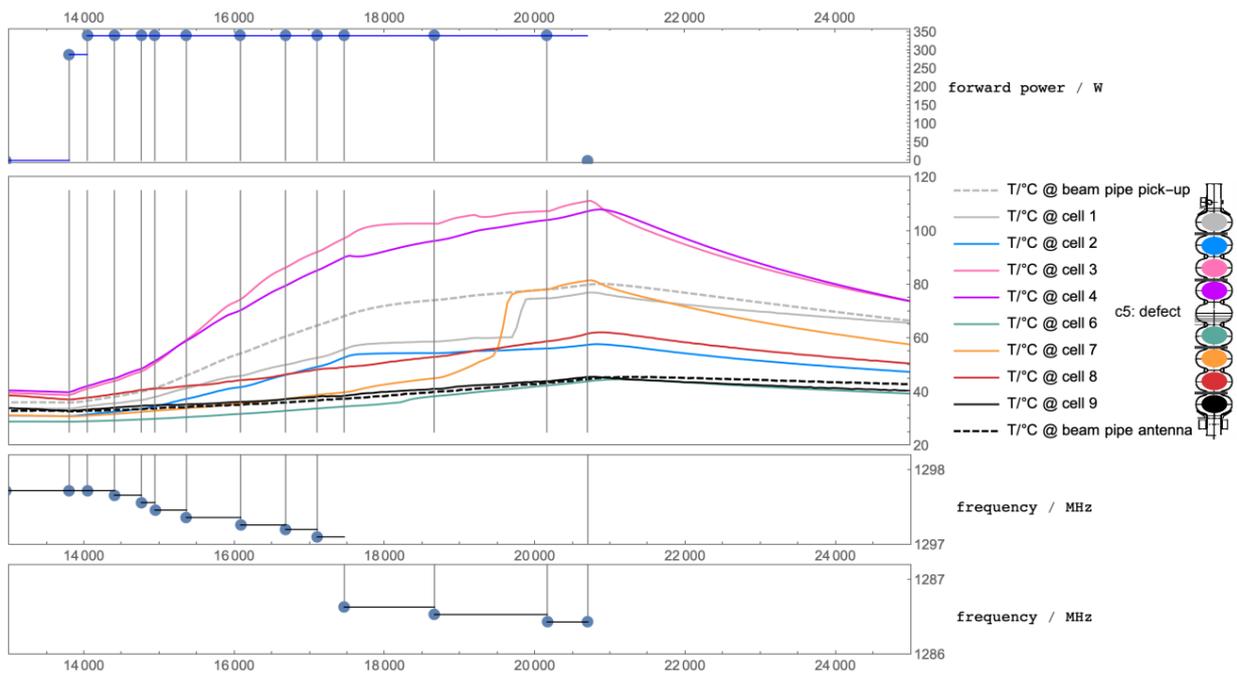

Figure 3: Continuation of the heating experiment shown in Fig. 2, legends similar except of omission of the transmitted power signal and a split frequency display. The latter was chosen in order to give a proper fine resolution even though the frequency was changed from the π-mode ($t < 17.5 \cdot 10^3$ s, upper frequency graph) to the 5π/9-mode ($t > 17.5 \cdot 10^3$ s, lower frequency graph).

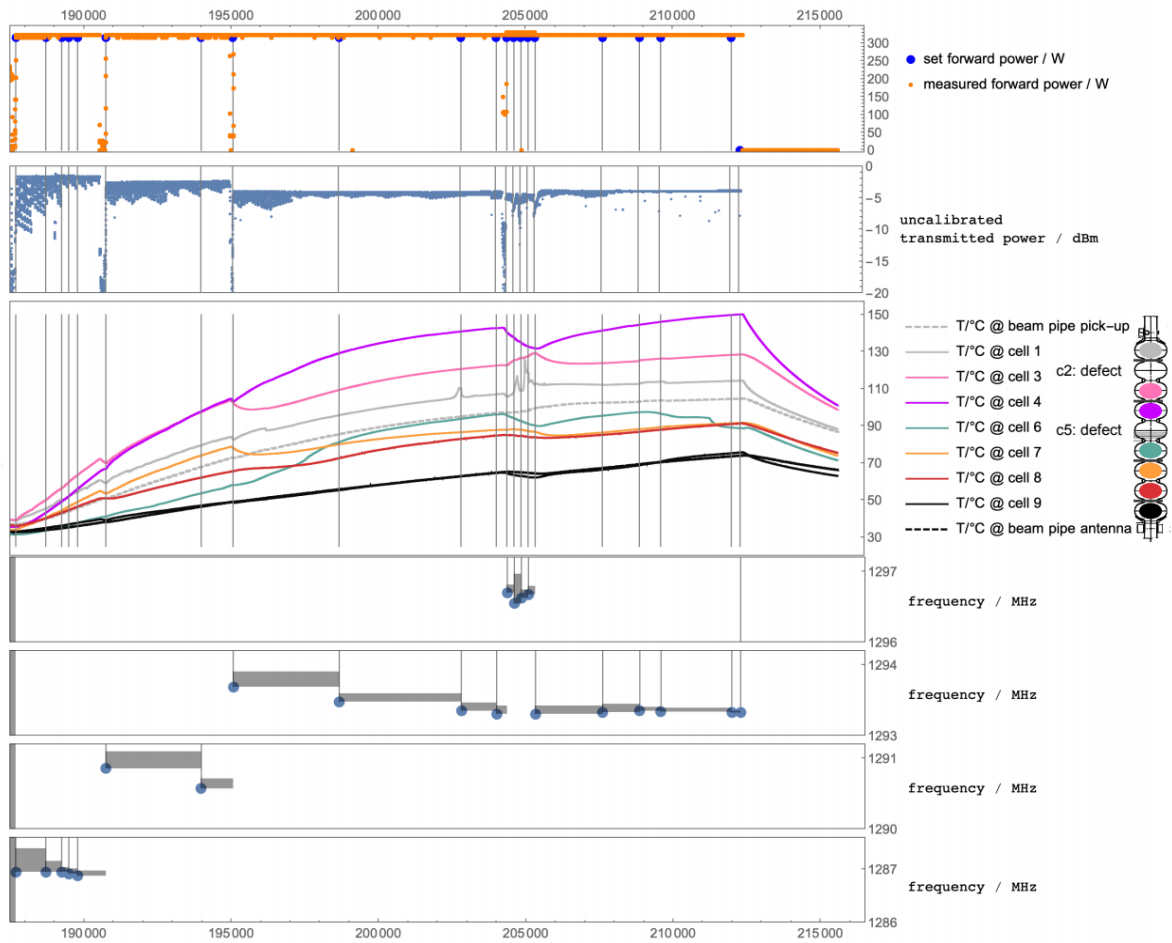

Figure 4: Heating experiment using four different fundamental passband modes, the amplifier driven by a VNA with narrow-band frequency sweeps (all graphs depending on the same time scale, given in seconds, blue dots and vertical lines indicating change of set values). *Bottom to top:* VNA frequency band settings (MHz, grey blocks indicating sweep width) in the $5\pi/9$-, $6\pi/9$-, $7\pi/9$- and $\pi$-mode; temperature measurements in all cells (except of cells 2 and 5, suffering from sensor defects) and both beam pipe extensions (°C); (uncalibrated) measured pick-up power (dBm); amplifier forward power set values (W, blue) and readout values (orange) showing a few short power trips.

## IV. CONCLUSIONS AND OUTLOOK

We tested the rf-power driven heating of a 9-cell 1.3 GHz accelerator cavity in order to gain early operational experiences with such a process and to derive indications for a more elaborated set-up to come. This then should be capable to reach a performance – i.e. temperature gain and flatness, heating rate and process control – appropriate for performance-improving in-vacuum baking of multicell structures. This in particular means the restriction of radiative losses, the use of a less invasive coupler, then clearly demanding for higher amplifier power, and the control of the temperature- and frequency-dependent field profile. Whether the severely inhomogeneous heating rates, which we observed here, will be possible to overcome with those means, will be one of the central objectives of such an improved experimental setting, which is currently under preparation. A later real-world application will not be able to rely on individual temperature sensors on every cell (nor on additional field profile data), but will need to derive such data from a proper model, transferring rf transmission data into temperature-dependent cell dimensions. The upcoming experiments shall serve also to develop and test such a model. In case of success, the sequence of cold testing, rf-powered heat treatment and second cold testing without intermediate venting or dismantling should be demonstrated.

# REFERENCES

[1] Y. Tamashevich, A. Prudnikava, A. Matveenko, J. Knobloch, "Moderate temperature heat treatment of a niobium superconducting cavity without oven", Helmholtz-Zentrum Berlin SRF Note 2023-06, https://doi.org/10.48550/arXiv.2307.09094

[2] G. Wu et al., Medium temperature furnance baking of low-beta 650 MHz five-cell cavities, 21th Int. Conf.RF Supercond., SRF2023, Grand Rapids, MI, USA, `doi:10.18429/JACoW-SRF2023-MOPMB030`

[3] B. Aune et al., "Superconducting TESLA cavities" Phys.Rev.STAB, Vol. 3, 092001, 2000, https://doi.org/10.1103/PhysRevSTAB.3.092001

[4] I. Sanz Troyano, "Design and Development of a RF-Based Conditioning for Performance Recovery on Superconducting (SRF) Cavities", Master Thesis Universidad Politécnica de Madrid, June 26, 2024.

[5] J. Knobloch et al., "Status of the HoBiCaT Superconducting Cavity Test Facility at BESSY", in Proc. 9th European Particle Accelerator Conf. EPAC'04, Lucerne, Switzerland, 5.-9.July 2004. TUPKF008, pp. 970-972.